\documentclass[twocolumn,showpacs,aps,superscriptaddress]{revtex4-1}
\usepackage{amsmath}
\usepackage{graphicx}
\usepackage{dcolumn}
\usepackage{float}
\usepackage{graphicx}
\usepackage{epstopdf}
\usepackage{epsfig}

\usepackage[colorinlistoftodos,prependcaption]{todonotes}

\begin{document}
\title{Investigation of the magnetic structure and crystal field states of pyrochlore antiferromagnet Nd$_2$Zr$_2$O$_7$}
\author{J. Xu}
\altaffiliation{jianhui.xu@helmholtz-berlin.de}
\affiliation{\mbox{Helmholtz-Zentrum Berlin f\"{u}r Materialien und Energie GmbH, Hahn-Meitner Platz 1, D-14109 Berlin, Germany}}
\affiliation{\mbox{Institut f\"{u}r Festk\"{o}rperphysik, Technische Universit\"{a}t Berlin, Hardenbergstraße 36, D-10623 Berlin, Germany}}
\author{V. K. Anand}
\altaffiliation{vivekkranand@gmail.com}
\affiliation{\mbox{Helmholtz-Zentrum Berlin f\"{u}r Materialien und Energie GmbH, Hahn-Meitner Platz 1, D-14109 Berlin, Germany}}
\author{A. K. Bera}
\affiliation{\mbox{Helmholtz-Zentrum Berlin f\"{u}r Materialien und Energie GmbH, Hahn-Meitner Platz 1, D-14109 Berlin, Germany}}
\affiliation{Solid State Physics Division, Bhabha Atomic Research Centre, Mumbai 400085, India}
\author{M. Frontzek}
\affiliation{\mbox{Paul Scherrer Institute, 5232 Villigen PSI, Switzerland}}
\author{D. L. Abernathy}
\affiliation{Instrument and Source Division, Neutron Sciences Directorate, Oak Ridge National Laboratory, Oak Ridge,
Tennessee 37831, USA}
\author{N. Casati}
\affiliation{\mbox{Paul Scherrer Institute, 5232 Villigen PSI, Switzerland}}
\author{K. Siemensmeyer}
\affiliation{\mbox{Helmholtz-Zentrum Berlin f\"{u}r Materialien und Energie GmbH, Hahn-Meitner Platz 1, D-14109 Berlin, Germany}}
\author{B. Lake}
\altaffiliation{bella.lake@helmholtz-berlin.de}
\affiliation{\mbox{Helmholtz-Zentrum Berlin f\"{u}r Materialien und Energie GmbH, Hahn-Meitner Platz 1, D-14109 Berlin, Germany}}
\affiliation{\mbox{Institut f\"{u}r Festk\"{o}rperphysik, Technische Universit\"{a}t Berlin, Hardenbergstraße 36, D-10623 Berlin, Germany}}

\date{\today}

\begin{abstract}
We present synchrotron x-ray diffraction, neutron powder diffraction and time-of-flight inelastic neutron scattering measurements on the rare earth pyrochlore oxide Nd$_2$Zr$_2$O$_7$ to study the ordered state magnetic structure and cystal field states. The structural characterization by high-resolution synchrotron x-ray diffraction confirms that the pyrochlore structure has no detectable O vacancies or Nd/Zr site mixing. The neutron diffraction reveals long range all-in/all-out antiferromagnetic order below $T_{\rm N}\approx 0.4$~K with propagation vector {\bf k} = (0 0 0) and an ordered moment of $1.26(2)\,\mu_{\rm B}$/Nd at 0.1~K. The ordered moment is much smaller than the estimated moment of $2.65\,\mu_{\rm B}$/Nd for the local $\langle 111\rangle$ Ising ground state of Nd$^{3+}$ ($J=9/2$) suggesting that the ordering is partially suppressed by quantum fluctuations. The strong Ising anisotropy is further confirmed by the inelastic neutron scattering data which reveals a well-isolated dipolar-octupolar type Kramers doublet ground state. The crystal field level scheme and ground state wavefunction have been determined.
\end{abstract}


\maketitle

\section{\label{Intro} INTRODUCTION}

Frustrated magnetism in pyrochlore oxides is undergoing intense investigation due to their emergent novel magnetic ground states and excitations arising from competing interactions \cite{book,rev2010}. Over the past two decades many interesting and exotic magnetic and thermodynamic phenomena have been observed in the rare earth pyrochlore oxides $R_2B_2$O$_7$ ($R$ is a trivalent rare earth ion and $B$ a tetravalent transition metal ion or Ge, Sn, Pb) which contain magnetic networks of corner-sharing tetrahedra \cite{rev2010,pressure,pressure2,lead}. For example, Dy$_2$Ti$_2$O$_7$ and Ho$_2$Ti$_2$O$_7$ have the spin-ice ground state and their excitations are magnetic monopoles \cite{mononature,monotennant,monofennel}, Tb$_2$Ti$_2$O$_7$ has a spin-liquid ground state \cite{rev2010,tbtio} and Er$_2$Ti$_2$O$_7$ develops antiferromagnetic ordering through an order-by-disorder mechanism \cite{rev2010,ertio}. The nature of the ground state of the rare earth pyrochlores depends on three competing interactions: the exchange interaction, the dipolar interaction and the crystal electric field (CEF) \cite{rev2010}. Among these, the crystal field produced at the rare earth cations by the surrounding oxygen anions, is usually strongest and dominates much of the underlying physics. Accordingly, the $R_2B_2$O$_7$ compounds display strongly anisotropic magnetic behavior, e.g. local $\langle$111$\rangle$ Ising anisotropy in Dy$_2$Ti$_2$O$_7$ and $XY$-anisotropy (with a local $\langle$111$\rangle$ easy plane) in Er$_2$Ti$_2$O$_7$ \cite{rev2010}. Furthermore, the delicate balance between the exchange and dipolar interactions leads to a diverse phase diagram, such as in the Gd pyrochlores Gd$_2$Ti$_2$O$_7$ and Gd$_2$Sn$_2$O$_7$ for which anisotropy is very weak \cite{Palmer-Chalker,gto1,gto2,gto3}.

The rare earth pyrochlore oxides with Ising anisotropy present a very interesting phase diagram. Monte Carlo simulations have revealed that when the ferromagnetic (FM) dipolar interaction dominates over the antiferromagnetic (AFM) exchange interaction, the ground state is the spin-ice state (two spins pointing into and two spin pointing out of the tetrahedra - the so called `2-in/2-out configuration') \cite{dsim2000, dsim2004}. In contrast when the AFM exchange interaction dominates, an `all-in/all-out' (AIAO) antiferromagnetic order can be stabilized where the spins alternate between pointing all into and all out of successive tetrahedra \cite{dsim2000, dsim2004}. Several pyrochlores have been found to show the spin-ice state (e.g. Dy$_2$Ti$_2$O$_7$ and Ho$_2$Ti$_2$O$_7$) and the underlying physics of the spin-ice phase has been investigated extensively by various complementary techniques \cite{rev2010}. However, a systematic investigation of the AIAO AFM ordered phase is still lacking as only very few compounds showing this magnetic structure are known. Recent theoretical studies of the AIAO ordered phase suggest that this state can show the magnetic Coulomb phase like spin-ice systems whereby new fascinating concepts of `double monopoles' and `staggered charge fluid and crystal' have been introduced \cite{frangmen, Guruciaga2014}, making the study of the AIAO AFM ordering and associated dynamics very important.  

In order for the AFM AIAO magnetic structure to be stabilized, the ferromagnetic dipolar interactions must be weak so that the antiferromagnetic exchange interactions can dominate the physics. This is most likely to occur for the light rare earth ions which have smaller moments and therefore smaller dipole interactions. Nd$^{3+}$ is a promising candidate because it is among the lightest rare earth ions and usually has Ising anisotropy in the pyrochlore environment with the magnetic moments oriented preferentially along the local $\langle$111$\rangle$ direction in each tetrahedron. Furthermore Nd$^{3+}$ is a Kramers ion ($J=9/2$) and has also been predicted to be a `dipolar-octuplar' doublet  whereby two distinct quantum spin-ice (QSI) phases namely dipolar QSI and octupolar QSI are possible \cite{dioctu}. Thus Nd-based pyrochlores are very promising for the study of new exotic phenomena. AIAO AFM ordering has recently been found in the iridate pyrochlore  Nd$_2$Ir$_2$O$_7$ on the sublattices of both the Nd$^{3+}$ ion and the magnetic transition metal ion Ir$^{4+}$ \cite{ndiro1,ndiro2}. However the magnetic behavior of this compound is dominated by the much stronger exchange interactions between the Ir$^{4+}$ ions which control the  order of the Nd$^{3+}$ ions. In Nd$_2$Mo$_2$O$_7$ the long range ordering of Nd$^{3+}$ moments is found to be ferromagnetic with the two-in/two-out spin configuration \cite{Yasui2001, Yasui2003}, however again their ordering in this compound is strongly influenced by the presence of $d$-electron moments on the Mo$^{4+}$ ions. In our recent investigation, we found the AIAO AFM ordering of Nd$^{3+}$ in the hafnate pyrochlore, Nd$_2$Hf$_2$O$_7$ \cite{Anand2015}. In this compound the Hf$^{4+}$ is non-magnetic and the order is due entirely to the interactions between the Nd$^{3+}$ ions. Here, we extend our study on the related compound Nd$_2$Zr$_2$O$_7$ having non magnetic $B$ site ion to get further insight of the AIAO magnetic structure corresponding to the exotic dipolar excitation of doube monopoles.

Previous investigations of Nd$_2$Zr$_2$O$_7$ by Bl\"{o}te {\it et al}.\ \cite{Blote1969} report a peak at 0.37~K in heat capacity data which could be related to a possible magnetic phase transition to a long range ordered state. More recently a study of the magnetic susceptibility and heat capacity down to 0.5~K by Hatnean {\it et al}.\ \cite{nzoprb} reported a FM interaction between Nd$^{3+}$ moments (inferred from the positive Curie-Weiss temperature). A strong local $\langle$111$\rangle$ Ising anisotropy was also found and the susceptibility was used to deduce the crystal field parameters \cite{nzoprb}. While we were preparing this manuscript, a paper appeared on ArXiv \cite{nzoarxiv} which reports results for Nd$_2$Zr$_2$O$_7$ that are similar to ours including an Ising anisotropy and an AIAO AFM structure.

In this paper, we show that Nd$_2$Zr$_2$O$_7$ indeed has long range antiferromagnetic ordering below $T_{\rm N}\approx 0.4$~K\@. Our neutron diffraction (ND) data reveal an AIAO magnetic structure of the Nd$^{3+}$ moments with a propagation vector ${\bf k} = (0\,0\,0)$ and ordered moment $m = 1.26(2) \, \mu_{\rm B}$/Nd at 0.1~K\@. The Ising anisotropy of the Nd$^{3+}$ ion is confirmed by the analysis of magnetic susceptibility $\chi(T)$ and isothermal magnetization $M(H)$ data.  We have measured the crystal field excitations up to 400 meV by inelastic neutron scattering (INS) which reveal that the Kramers doublet ground state is well separated (by 23.4~meV) from the first excited state. We find that the ground state magnetic properties of Nd$_2$Zr$_2$O$_7$ are well described by an effective spin $S = 1/2$ with effective $g$-factor $g_{zz} = 5.30(6)$ and $g_{\bot}=0$. Accordingly a moment of $2.65\,\mu_{\rm B}$/Nd is expected for the Ising ground state. Contrary to such expectation the ordered state moment $1.26(2) \, \mu_{\rm B}$/Nd is much lower which suggests the presence of strong fluctuations even in the ordered state. Finally the ground state wavefunction is found to be compatible with a dipolar-octupolar doublet, making Nd$_2$Zr$_2$O$_7$ a candidate compound for dipolar and octupolar spin ice phases \cite{dioctu}.

\section{\label{ExpDetails} EXPERIMENTAL DETAILS}

Polycrystalline Nd$_2$Zr$_2$O$_7$ sample was synthesized by solid state route by firing the stoichiometric mixture of Nd$_2$O$_3$ (99.99\%) and ZrO$_2$ (99.99\%) in an alumina crucible at 1200, 1400, 1550~$^{\circ}$C in air for 8 days with several intermediate grindings and pelletizing. The nonmagnetic reference compound La$_2$Zr$_2$O$_7$ was also prepared by the same method and used to estimate the phonon contribution to the neutron inelastic scattering. The qualities of the samples were checked by room temperature powder x-ray diffraction (XRD) using the laboratory-based diffractometer (Bruker-D8, Cu-$K_\alpha$). The magnetic susceptibility and isothermal magnetization measurements were performed by using Quantum Design magnetic properties measurement system (MPMS) superconducting quantum interference device (SQUID) magnetometer and MPMS SQUID Vibrating Sample Magnetometer (VSM) at Mag Lab, Helmholtz-Zentrum Berlin (HZB), Germany. The heat capacity measurements were performed by using Quantum Design physical properties measurement system (PPMS), Mag Lab, HZB.

The high-resolution synchrotron x-ray diffraction patterns were collected on MS-beamline \cite{mspsi} at Paul Scherrer Institute (PSI), Switzerland and the patterns were refined by using the FullProf Suite \cite{fullprof}. In order to carry out these synchrotron x-ray measurements the Nd$_2$Zr$_2$O$_7$ sample was ground very finely and mixed with diamond powder ($\sim30\%$) to reduce x-ray absorption. The mixture was put into a thin capillary (0.3~mm in diameter) which rotated continuously during the measurement to reduce the effect of preferred orientation. The data were collected for a number of temperatures between 290~K and 5~K with x-rays of energy 25~keV\@. The synchrotron XRD patterns of standard Si and LaB$_6$ powders (NIST) were also recorded at room temperature under the same conditions to accurately determine the wavelength and the instrument profile parameters, respectively.

The powder neutron diffraction measurements were performed on the Cold Neutron Powder Diffractometer (DMC) at PSI, Switzerland. About 10~g Nd$_2$Zr$_2$O$_7$ powder was sealed in a cylindrical copper can (diameter 10~mm) with filling high pressure $^4$He gas for a better thermalization. A dilution refrigerator was used to achieve the lowest temperature of 0.1~K\@. Long wavelength neutrons (3.80~{\AA}) were used in combination with a $2\theta$ angle range of 5$^\circ$--90$^\circ$ to achieve a good resolution in the low-$Q$ region for the determination of the characteristic {\bf k} vector of magnetic structure. Shorter wavelength neutrons (2.46~{\AA}) were used over $10^\circ \leq 2\theta \leq 90^\circ$ to access a larger $Q$ space region for the magnetic structure refinement. Data at several temperatures between 0.1~K and 4~K were collected with a counting time of $\sim 5$ hours for every temperatures. The ND data were also refined by the FullProf Suite \cite{fullprof}.

The inelastic neutron scattering measurements were performed on the direct geometry time-of-flight spectrometer ARCS at the Spallation Neutron Source (SNS), Oak Ridge National Laboratory (ORNL), USA. The INS measurements were conducted on about 20~g samples of both Nd$_2$Zr$_2$O$_7$ and La$_2$Zr$_2$O$_7$.  Thin-walled cylindrical aluminium cans were used to mount the powdered samples. The INS data were recorded with incident neutron energies of $E_i = 50$~meV, 150~meV and 400~meV at 5~K and 300~K to access the full range of the excitations. The INS data were analyzed by the software SPECTRE \cite{SPECTRE}.

\section{\label{Structure} Crystallography}

\subsection{Laboratory x-ray diffraction}

The room temperature powder XRD data (not shown) of Nd$_2$Zr$_2$O$_7$ were refined using the FullProf software \cite{fullprof}. The structural Rietveld refinement revealed the sample to be single phase and confirmed the ${\rm Eu_2Zr_2O_7}$-type face-centered cubic (fcc) pyrochlore structure (space group $Fd\bar{3}m$) of Nd$_2$Zr$_2$O$_7$ with the lattice parameter $a = 10.6728(1)$ and the $x$-coordinate of O1 $x_{\rm O1} = 0.3351(4)$, which agrees very well with the reported values \cite{nzocp,nzoprb}. The single phase nature of the La$_2$Zr$_2$O$_7$ sample was also inferred from the refinement of the room temperature powder XRD data (not shown), which was also found to crystallize in the fcc pyrochlore structure with parameters $a = 10.7996(1)$~\AA\ and $x_{\rm O1} = 0.3310(7)$, again in very good agreement with the values in literature \cite{Pruneda2005}.

\subsection{\label{Structure} Synchrotron x-ray diffraction}

\begin{figure}
\includegraphics[width=3in, keepaspectratio]{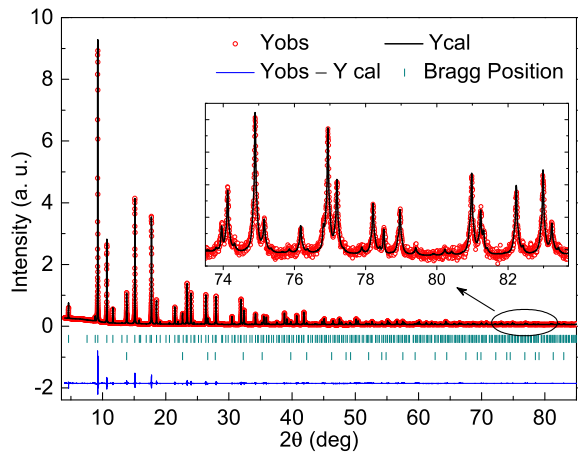}
\caption {(Color online) Synchrotron x-ray diffraction pattern (red circle) of polycrystalline Nd$_2$Zr$_2$O$_7$ at 5~K along with structural Rietveld refinement profile [simulated (black line) and difference (blue line)]. The short vertical bars indicate the positions of Bragg peaks of Nd$_2$Zr$_2$O$_7$ and diamond (added to the sample to reduce x-ray absorption). Inset: An expanded scale view showing the details of the refinement in a small angle range at high $2\theta$.}
\label{fig:XRD}
\end{figure}

\begin{figure}
\includegraphics[width=3in, keepaspectratio]{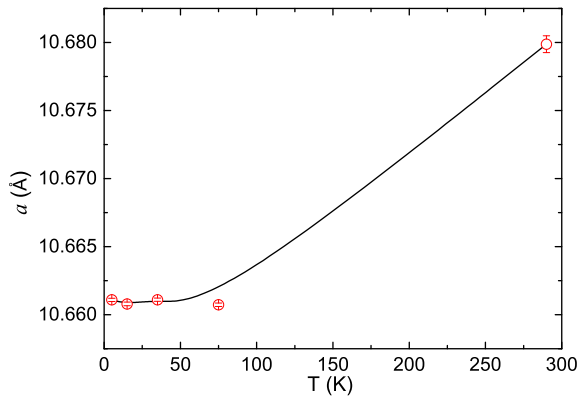}
\caption {(Color online) Temperature $T$ dependence of lattice parameter $a$. The black line is the guide to eyes.}
\label{fig:axT}
\end{figure}

Figure~\ref{fig:XRD} shows the high-resolution synchrotron powder XRD pattern recorded at 5.0~K together with the structural Rietveld refinement profile calculated for the fcc pyrochlore structure. The wavelength of the synchrotron x-rays used was determined to be 0.495734(8) {\AA} by refining the Si pattern and the starting profile parameters were obtained by refining the LaB$_6$ pattern collected with the same instrument settings. While refining we included the possibility of site mixing of Nd and Zr and allowed the occupancies to vary for possible off-stoichiometry. The crystallographic parameters and agreement factors obtained from the refinements of the 5~K and 290~K patterns are listed in Table~\ref{tab:XRD}. The refinements of the 5~K and 290~K patterns indicate no change in structural symmetry between the room temperature and 5~K\@. The temperature dependences of lattice parameter $a$ is shown in Fig.~\ref{fig:axT}. A weak contraction of the unit cell is inferred from the $T$ dependence of $a$. Furthermore no evidence of Nd/Zr site mixing or oxygen deficiency could be deduced from the refinement. 

\begin{table}
\caption{\label{tab:XRD} Crystallographic parameters for Nd$_2$Zr$_2$O$_7$ obtained from the refinements of high-resolution synchrotron x-ray diffraction patterns at 5~K and 290~K\@. The Wyckoff positions of Nd, Zr, O1 and O2 atoms in space group $Fd\bar{3}m$ are 16$d$ (1/2,1/2,1/2), 16$c$ (0,0,0), 48$f$ ($x_{\rm O1}$,1/8,1/8) and 8$b$ (3/8,3/8,3/8), respectively. The atomic coordinate $x_{\rm O1}$ is listed below. }
\begin{ruledtabular}
\begin{tabular}{lcc}
 &  5 K & 290 K \\
 \hline
 \underline{Lattice parameters}\\
{\hspace{0.8cm} $a$ ({\AA})} &  10.6611(1) & 10.6735(7) \\	
\underline{Atomic coordinate}\\
\hspace{0.8cm} $x_{\rm O1}$ & 0.3357(2) & 0.3356(2)\\
\underline{Refinement quality} \\
\hspace{0.8cm} $\chi^2$   & 10.3 & 8.35 \\	
\hspace{0.8cm} $R_{\rm p}$ (\%)  & 3.01 & 4.05\\
\hspace{0.8cm} $R_{\rm wp}$ (\%) & 3.73 & 4.78\\
\hspace{0.8cm} $R_{\rm Bragg}$ (\%) & 4.12 & 4.57\\
\end{tabular}
\end{ruledtabular}
\end{table}

The $R_2B_2$O$_7$ pyrochlore structure can be viewed as an ordered defect fluorite CaF$_2$ structure (space group $Fm\bar{3}m$) \cite{rev1983}. In fluorite structure, Ca cations form a cubic close packed lattice and fluorine anions fill all the tetragonal interstices. In the pyrochlore structure, there are two types of $R$-$B$ ordered close packed layers stacked alternatively along the [111] direction: one with Kagom\'{e} lattice formed by $R$ atoms with $B$ atoms located at the hexagon centers, and the other with the reversed $R/B$ occupation. The O$^{2-}$ anions sit in the tetrahedral interstices formed by the $R_4$ (8$b$ site) and the $R_2B_2$ (48$f$ site) networks, leaving the $B_4$ interstices (8$a$ site) unoccupied. The ratio of the ionic radii of the trivalent ($r_R$) and tetravalent ($r_B$) cations determines whether the ordered or disordered phase forms. At ambient pressure, a stable ordered pyrochlore phase is found for $1.36~<r_R/r_B<1.71$ \cite{rev2010}. For Nd$_2$Zr$_2$O$_7$, the effective ionic radii of eight-fold coordinated Nd$^{3+}$  and six-fold coordinated Zr$^{4+}$ are 1.109 {\AA} and 0.72 {\AA}, respectively \cite{radii}. Thus $r_R/r_B = 1.54$ which indicates that Nd$_2$Zr$_2$O$_7$ lies well inside the stable pyrochlore phase. In addition, the valence of Zr$^{3+}$ is very rare in Zr compounds, thus O deficiency in zirconate pyrochlores should not appear as commonly as in the titanate pyrochlores where Ti$^{3+}$ is possible, especially after heating in air \cite{ovacancy}. Altogether this suggests that Nd$_2$Zr$_2$O$_7$ should form a highly ordered pyrochlore phase as is inferred from the synchrotron x-ray data.

\section{\label{Magnetic} Magnetic Susceptibility and Magnetization}

\begin{figure}
\includegraphics[width=3in, keepaspectratio]{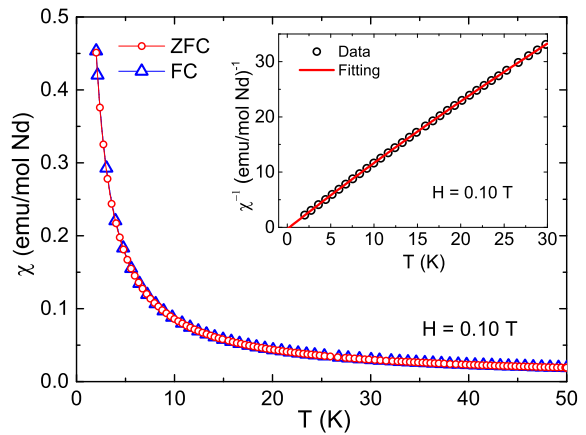}
\caption {(Color online) Temperature $T$ dependence of susceptibility $\chi$ of polycrystalline Nd$_2$Zr$_2$O$_7$ for 2~K~$\leq T\leq 50$~K measured in an applied field $H=0.10$~T\@. Inset: Inverse susceptibility $\chi^{-1}(T)$ for 2~K~$\leq T\leq 30$~K at $H = 0.10$~T and the Curie-Weiss fitting (red solid line) in 10~K~$\leq T\leq 30$~K\@.}
\label{fig:mag1}
\end{figure}

Figure~\ref{fig:mag1} shows the zero field cooled (ZFC) and field cooled (FC) magnetic susceptibility $\chi$ of Nd$_2$Zr$_2$O$_7$ as a function of temperature $T$ for 2~K~$\leq T\leq 50$~K measured in a field of $H=0.10$~T after subtracting the diamagnetic signal of the sample holder. The $\chi$ increases smoothly with decreasing $T$ and shows no anomaly or thermal hysteresis between ZFC and FC data above 2~K, in agreement with the single crystal data \cite{nzoprb}. The $\chi(T)$ data were fitted to a modified Curie-Weiss law $\chi(T) = \chi_0 + C/(T-\theta_{\rm p})$ in the temperature range 10~K~$\leq T\leq 30$~K to minimize the effects of short range magnetic correlations (at $T<10$~K) and crystal field excitations (at high-$T$) on the estimate of the Weiss temperature $\theta_{\rm p}$ and effective moment $\mu_{\rm eff}$ of the ground state. The best fit (inset of Fig.~\ref{fig:mag1}) gives $\chi_0 = 2.82(2) \times 10^{-3}$~emu/mol\,Nd,  $\theta_{\rm p} = 0.233(5)$~K and $\mu_{\rm eff} = 2.55(1) \,\mu_{\rm B}$/Nd. In order to compare the values of $\theta_{\rm p}$ and $\mu_{\rm eff}$ with those reported for a single crystal Nd$_2$Zr$_2$O$_7$ \cite{nzoprb}, we also fitted the susceptibility by $\chi(T) = C/(T-\theta_{\rm p})$ for 2~K~$\leq T\leq 10$~K which gives $\theta_{\rm p}= 0.124(2)$~K and $\mu_{\rm eff} = 2.60(1)\, \mu_{\rm B}$/Nd, consistent with the reported values. The positive $\theta_{\rm p}$ indicates an effective ferromagnetic interaction between the Nd$^{3+}$ spins. This contrasts with the antiferromagnetically ordered ground state inferred from neutron diffraction data as shown later. The $\mu_{\rm eff} = 2.55(1)\,\mu_{\rm B}$/Nd obtained is much lower than the theoretical paramagnetic state moment of $3.62 \,\mu_{\rm B}$/Nd for free Nd$^{3+}$ ions ($\mu_{\rm eff}= g_J \sqrt{J(J+1)}$\,) and reflects that the crystal field ground state is not the pure $|^4I_{9/2},\pm 9/2 \rangle$ Kramers doublet.

\begin{figure}
\includegraphics[width=3in, keepaspectratio]{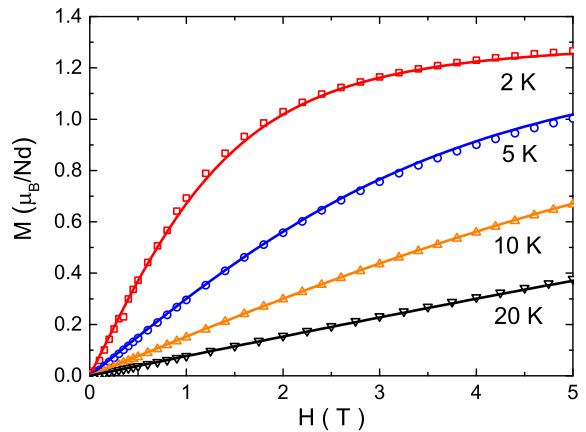}
\caption {(Color online) Magnetic field $H$ dependence of isothermal magnetization $M$ of polycrystalline Nd$_2$Zr$_2$O$_7$ at 2~K, 5~K, 10~K and 20~K\@. The solid curves are the fits based on the Ising anisotropic model according to Eq.~(\ref{MH-Ising}).}
\label{fig:mag2}
\end{figure}

Figure~\ref{fig:mag2} shows the isothermal magnetization $M(H)$ curves at 2~K, 5~K, 10~K and 20~K\@. The $M(H)$ at 2~K shows a saturation tendency with $M= 1.27\,\mu_{\rm B}$/Nd at 5~T which is much lower than the free ion saturation value of $M_s=  g_J J\,\mu_{\rm B} = 3.27 \,\mu_{\rm B}$/Nd. The strongly reduced value of $M_s$ can be attributed to the strong Ising anisotropy and the reduction of the Nd$^{3+}$ moment due to the CEF effect (Sec.~\ref{INS}). To estimate the moment of Nd$^{3+}$ in Nd$_2$Zr$_2$O$_7$, we analyzed the $M(H)$ data with the effective spin-half model. For an Ising pyrochlore with large separation between the ground state doublet and the first excited state, which is the case with the present compound, the low temperature magnetic properties can be described by an effective spin $S = 1/2$ and the powder averaged magnetization in paramagnetic state can be described by \cite{isingmh}
\begin{equation}
\langle M \rangle = \frac{(k_{\rm B} T)^2}{g_{zz}\mu_{\rm B} H^2 S} \int_0^{g_{zz}\mu_{\rm B} H S/k_{\rm B} T} x \tanh(x) \,dx
\label{MH-Ising}
\end{equation}
where $x = g_{\rm zz}\mu_{\rm B} H S/k_{\rm B} T$ and $g_{zz}$ is the longitudinal $g$-factor (while the transverse $g$-factor is zero). Simultaneous fitting of $M(H)$ data at 2~K, 5~K, 10~K and 20~K (Fig.~\ref{fig:mag2}) yields $g_{\rm zz}= 5.24(2)$. This $g$-factor is lower than the $g_{zz} = 2 g_J J  = 6.54 $ expected for a Kramers doublet formed from only the pure $m_J = \pm 9/2$ states of Nd$^{3+}$, and suggests mixing of the $m_J$ states as is found from the CEF analysis of INS data below (Sec.~\ref{INS}). The $g_{zz}$ value yields the ground state moment of $m_{\rm Nd}=g_{\rm zz}S\,\mu_{\rm B} \approx 2.62 \,\mu_{\rm B}/$Nd for Nd$_2$Zr$_2$O$_7$ which is consistent with the $\mu_{\rm eff}$ determined from the susceptibility above. The obtained $g_{zz}$ value is comparable with that of Nd$_2$Hf$_2$O$_7$ for which a similar analysis of magnetic data has been found to yield $g_{zz} = 5.01(3)$ \cite{Anand2015}.

\section{\label{HeatCapacity} Heat Capacity}

\begin{figure}
\includegraphics[width=3in, keepaspectratio]{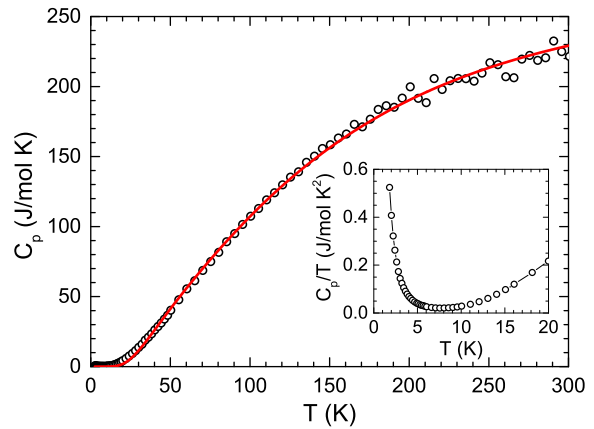}
\caption {(Color online) Temperature $T$ dependence of heat capacity $C_{\rm p}$ of Nd$_2$Zr$_2$O$_7$ for 1.8~K~$\leq T \leq$~300~K measured in zero field. The solid curve is the fit by Debye+Einstein models of lattice heat capacity plus crystal field contribution. Inset: $C_{\rm p}/T$ versus $T$ plot for $1.8~{\rm K} \leq T \leq 20$~K\@.}
\label{fig:HC}
\end{figure}

Figure~\ref{fig:HC} shows the heat capacity $C_{\rm p}(T)$ data for the temperature range 2~K--300~K measured in zero field. Consistent with the $\chi(T)$ data, the $C_{\rm p}(T)$ shows no anomaly related to a magnetic phase transition down to 2~K. However, an upturn is evident at $T<6$~K as can be seen from the $C_{\rm p}/T$ versus $T$ plot shown in the inset of Fig.~\ref{fig:HC}. The upturn in $C_{\rm p}(T)$ probably reflects the presence of short-range magnetic correlation at temperatures which is much higher than the antiferromagnetic ordering temperature of 0.4~K (Sec.~\ref{ND}). The upturn feature in  $C_{\rm p}(T)$ is consistent with the reported Lambda-type anomaly at 0.37~K \cite{Blote1969,nzocp} on account of long-range antiferromagnetic ordering. 

The low-$T$ $C_{\rm p}(T)$ data were fitted to $C_{\rm p}(T) = \gamma T + \beta T^{3} + \delta T^{5}$ in the temperature range $9.5~{\rm K} \leq T \leq 16$~K (with $\gamma = 0$ for an insulating ground state). The fit yields $\beta= 1.57(6) \times 10^{-4}$~J/mole\,K$^{4}$ and $\delta = 1.20(3) \times 10^{-6}$~J/mole\,K$^{6}$. The coefficient $\beta$ gives Debye temperature $\Theta_{\rm D} = 514(6)$~K according to the relation $\Theta_{\rm D} = (12 \pi^{4} n R /{5 \beta} )^{1/3}$ where $n=11$ is the number of atoms per formula unit and $R$ is the molar gas constant. However, the low-$T$ $C_{\rm p}(T)$ data in pyrochlores have been found to yield $\Theta_{\rm D}$ much smaller than the one obtained from high-$T$ $C_{\rm p}(T)$ data \cite{cpfit,Anand2015}. Therefore the $C_{\rm p}(T)$ data were analyzed by the combined Debye and Einstein models of lattice heat capacity \cite{cpfit}. For this purpose we included the crystal field contribution to the heat capacity $C_{\rm CEF}$ calculated according to CEF level scheme obtained from the analysis of INS data (Sec.~\ref{INS}). The sum of Debye+Einstein models of lattice heat capacity and $C_{\rm CEF}$ is shown by the solid curve in Fig.~\ref{fig:HC}. By fitting the difference $C_{\rm p}(T) - C_{\rm CEF}(T)$ data in the temperature range $10~{\rm K} \leq T \leq 300$~K we obtain $\Theta_{\rm D} = 756(10)$~K and Einstein temperature $\Theta_{\rm E} = 161(5)$~K with 73\% weight to the Debye term and 27\% to the Einstein term.  The deduced value of $\Theta_{\rm D}$ is very close to the one obtained for the similar compounds, such as for Nd$_2$Hf$_2$O$_7$ ($\Theta_{\rm D} = 785(6)$~K) \cite{Anand2015} and Dy$_2$Ti$_2$O$_7$ ($\Theta_{\rm D} = 722(8)$~K) \cite{cpfit}.

\section{\label{ND} Neutron Diffraction and Magnetic Structure}

\begin{figure}
\includegraphics[width=3in, keepaspectratio]{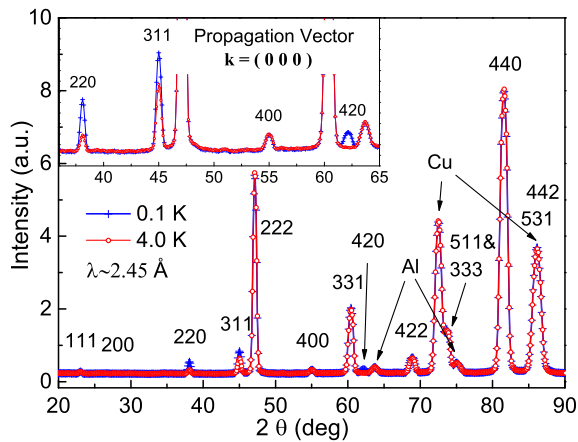}
\caption {(Color online) Comparison of the neutron diffraction patterns of Nd$_2$Zr$_2$O$_7$ collected at 0.1~K (blue) and 4.0~K (red). The peaks are marked with the ($hk\ell$) Miller indices. The Cu and Al peaks come from the sample holder and sample environment.}
\label{fig:nd1}
\end{figure}

\begin{figure}
\includegraphics[width=3.1in, keepaspectratio]{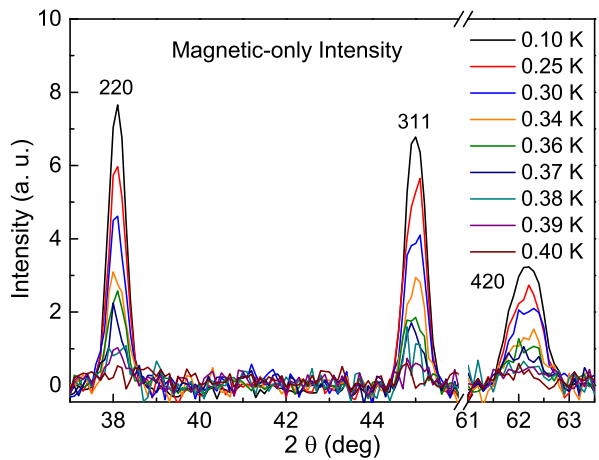}
\caption {(Color online) Temperature $T$ dependence of the intensities of the magnetic Bragg peaks (220) , (311) and (420) obtained by subtracting the 4.0~K neutron diffraction pattern from the neutron diffraction patterns at the indicated temperatures.}
\label{fig:nd20}
\end{figure}

The neutron diffraction data were collected at several temperatures between  0.1~K and 4.0~K\@. The ND patterns collected at 0.1~K and 4.0~K using neutrons of wavelength 2.45~\AA\ are shown in Fig.~\ref{fig:nd1}. As shown in the inset of Fig.~\ref{fig:nd1} additional intensities are clearly seen at 0.1~K, in particular on top of the nuclear Brag peaks (220) and (311) which indicates long range magnetic order. Further we see an additional magnetic peak (420) where nuclear Bragg reflection is forbidden, which further confirms the magnetic origin of the additional intensities. The magnetic intensity decreases continuously with increasing temperature (indicating a second order phase transition) and vanishes at 0.4~K (Fig.~\ref{fig:nd20}), allowing us to define $T_{\rm N} \approx 0.4$~K consistent with the reported heat capacity data \cite{Blote1969,nzocp} and the ND data in Ref. \cite{nzoarxiv}. No extra peaks were observed at low $Q$ in the ND pattern collected with neutrons of wavelength 3.80~\AA\ (not shown). 

\begin{figure}
\includegraphics[width=3in, keepaspectratio]{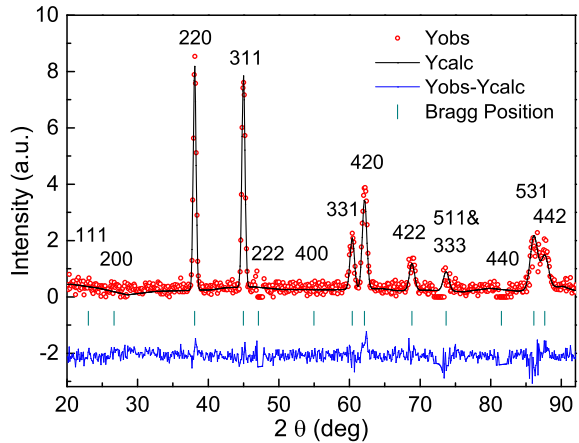}
\caption {(Color online) Magnetic diffraction pattern (red circles) at 0.1~K ( obtained from subtracting 4.0~K pattern from the 0.1~K pattern) together with the calculated magnetic refinement pattern (black line) for an ``all-in/all-out" magnetic structure. The difference between the experimental and calculated intensities is shown by the blue curve at the bottom. The green vertical bars show the magnetic Bragg peak positions. The peaks are marked with the ($hk\ell$) Miller indices.}
\label{fig:nd3}
\end{figure}

The difference pattern obtained by subtracting the 4.0~K ND data from the 0.1~K data is shown in Fig.~\ref{fig:nd3} which clearly shows all the magnetic Bragg peaks associated with the ordering of Nd$^{3+}$ moments. We found that all the magnetic Bragg peaks can be indexed with the magnetic propagation vector ${\bf k} = (0\,0\,0)$, which is consistent with Ref.~\cite{nzoarxiv}. The wavevector ${\bf k} = (0\,0\,0)$ also indexes all the magnetic Bragg peaks in Nd$_2$Hf$_2$O$_7$ \cite{Anand2015} indicating a similar magnetic structure. The representation analysis performed using the program BASIREPS shows that the magnetic representation of Nd (16$d$ site) can be reduced into four nonzero irreducible representations (IRs) of the little group of wavevector ${\bf k} = (0\,0\,0)$: 
\begin{equation}
\Gamma_{\rm mag\,Nd} = 1\,\Gamma_3^1 + 1\,\Gamma_6^2 + 1\,\Gamma_8^3 + 2\, \Gamma_{10}^3.
\label{eq:IRs}
\end{equation}
Each IR is multiplied by the number of times it occurs, the superscript of $\Gamma$ corresponds to the dimensionality of IR, and the subscript the order of IR. All possible models of the magnetic structure can be obtained by combinations of the basis vectors of the IRs \cite{Anand2015}.

\begin{figure}
\includegraphics[width=2.5in, keepaspectratio]{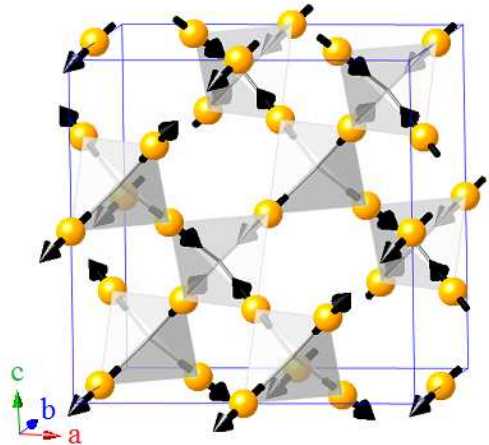}
\caption {(Color online) The `all-in/all-out' magnetic structure which is comprised of corner-shared tetrahedra with magnetic moments pointing alternatively inwards (all-in) or outwards (all-out) the centers of the successive tetrahedra (along the local $\langle$111$\rangle$ direction). The spheres represent Nd atoms and arrows denote the ordered moment directions. }
\label{fig:magstru}
\end{figure}

\begin{figure}
\includegraphics[width=3.1in, keepaspectratio]{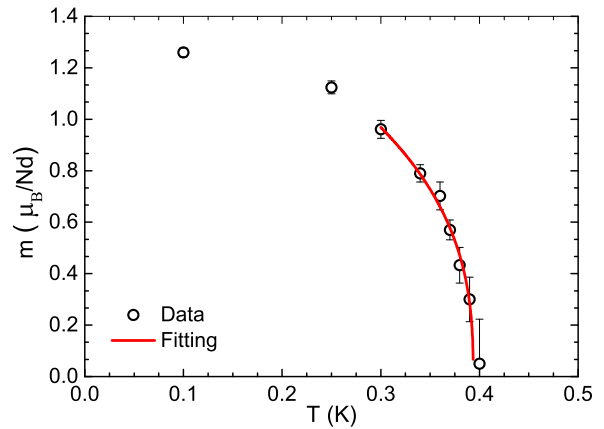}
\caption {(Color online) Temperature $T$ dependence of the ordered Nd$^{3+}$ moment $m(T)$ obtained from the refinement of neutron powder diffraction patterns at different temperatures. The solid curve shows the fitting in 0.3~K~$\leq T \leq T_{\rm N}$ by $m = m_0 (1 - T/T_{\rm N})^{\beta}$.}
\label{fig:nd21}
\end{figure}

We refined the ND difference pattern (magnetic only) with all possible magnetic structure models defined by the IRs. The crystallographic parameters were fixed to the ones determined in the synchrotron-XRD refinement and the scale factor was fixed by the refinement of nuclear pattern at 4.0~K\@. The best fit was obtained for the $\Gamma_3$ corresponding to the all-in/all-out magnetic structure shown in Fig.~\ref{fig:magstru}. The refinement of the magnetic-only pattern at 0.1~K is shown in Fig.~\ref{fig:nd3}. The ordered moment obtained from the refinement of ND data at 0.1~K is $m =1.26(2)\, \mu_{\rm B}$/Nd. The temperature dependence of ordered state moment is shown in Fig.~\ref{fig:nd21} and it is fitted to $m = m_0 (1 - T/T_{\rm N})^{\beta}$ giving $T_{\rm N}= 0.39(2)$~K and $\alpha=0.37(5)$ which within the error bar is close to the expected critical exponent 0.33 for a three dimensional Ising system. The same magnetic structure was found by Lhotel {\it et al}.\ \cite{nzoarxiv} who also found a similar $T_{\rm N}=0.41$~K for the powder sample but a much smaller ordered moment $0.80(5)\,\mu_{\rm B}$/Nd at 0.15~K. The difference is believed to be related to the synthesis procedure and quality of the sample. An ordered moment of $0.62(1)\, \mu_{\rm B}$/Nd at 0.1~K was found in Nd$_2$Hf$_2$O$_7$ which also orders with an AIAO AFM structure below 0.55~K\@ \cite{Anand2015}. The strong reduction in ordered moments in these Nd compounds reflect persistence of strong quantum fluctuations well inside the ordered state. We suspect that these reductions/fluctuations result from the octupolar component.

Having determined the $T_{\rm N}$ and magnetic structure of Nd$_2$Zr$_2$O$_7$, we now try to estimate the effective nearest neighboring dipolar interaction $D_{\rm nn}$ and effective exchange interaction $J_{\rm nn}$ to find out the position of Nd$_2$Zr$_2$O$_7$ in the phase diagram of Ising pyrochlore magnets \cite{dsim2000,dsim2004}. The $D_{\rm nn}$ can be simply estimated by using the equation \cite{rev2010} 
\begin{equation}
D_{\rm nn} = \frac{5}{3}\left(\frac{\mu_0}{4\pi}\right) \frac{m_{\rm Nd}^2}{r_{\rm nn}^3}.
\label{eq:Dnn}
\end{equation}
From the distance $r_{\rm nn} = (a/4)\sqrt 2 =3.77$~{\AA} between the nearest neighboring Nd$^{3+}$ ions and the ground state moment $m_{\rm Nd} = 2.65 \, \mu_{\rm B}$ (from the CEF analysis of the INS data, see below) we calculate the dipole interaction $D_{\rm nn}\approx 0.14$~K\@. With this we get $T_{\rm N}/D_{\rm nn} = 2.86$. According to the dipolar model phase diagram \cite{dsim2004}, this value of $T_{\rm N}/D_{\rm nn}$ places Nd$_2$Zr$_2$O$_7$ deep inside the AIAO AFM ordered phase with magnetic ordering wavevector ${\bf k} = (0\,0\,0)$ just as we have found experimentally. Furthermore the phase diagram suggests that for this value of $T_{\rm N}/D_{\rm nn}$, the ratio $J_{\rm nn}/D_{\rm nn} \approx -1.45$ giving the effective nearest neighbour exchange constant $J_{\rm nn}\approx-0.25$~K and thus the total interaction $J=J_{nn}+D_{nn}\approx-0.11$~K which is antiferromagnetic in contrast to the effective ferromagnetic interaction inferred from $\theta_{\rm p}$. We also see that $J_{\rm nn}$ is significantly stronger than $D_{\rm nn}$ which contrasts with the observation in spin-ice materials where the dipolar interaction is the stronger one. As such Nd$_2$Zr$_2$O$_7$ is expected not to show the frustration prevalent in spin-ice materials, but rather should result in a stable long-range ordered ground state.

\section{\label{INS} Inelastic Neutron Scattering and Crystal Field Excitations}

\begin{figure}
\includegraphics[width=3.6in, keepaspectratio]{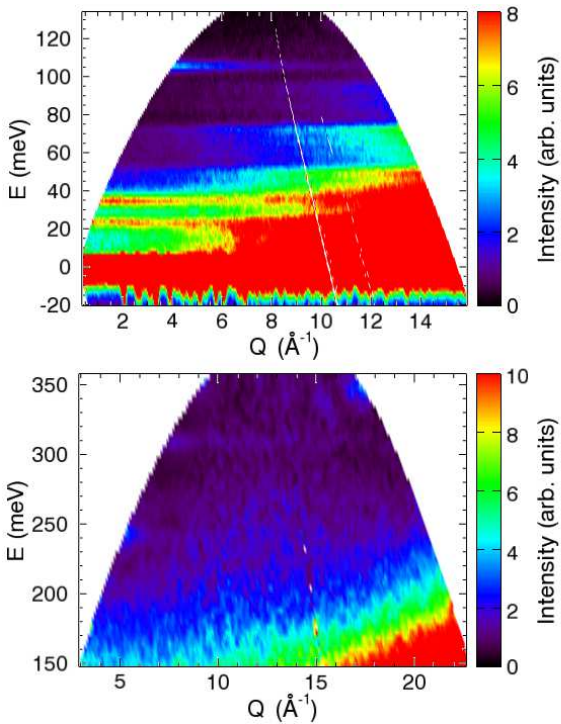}
\caption {(color online) Color contour maps of energy $E$ versus wavevector $Q$ of the inelastic neutron scattering intensity of Nd$_2$Zr$_2$O$_7$ at 5~K\@ with (a) $E_i=150$~meV and (b) $E_i=400$~meV. The spectra (a) shows the crystal field splitting of the ground state multiplet $^4I_{9/2}$ and (b) shows the crystal field splitting of the first excited multiplet $^4I_{11/2}$.}
\label{fig:cf1}
\end{figure}

\begin{figure}
\includegraphics[width=3in, keepaspectratio]{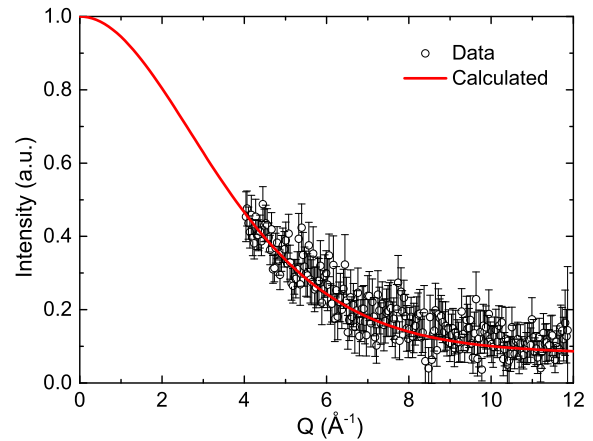}
\caption {(color online) $Q$ dependence of the integrated scattering intensity over energy region 105.7--106.7~meV. The solid curve is the scaled magnetic form factor $F^2(Q)$ with dipolar approximation  of Nd$^{3+}$.}
\label{fig:cfform}
\end{figure}

Figure~\ref{fig:cf1} shows the color contour maps of time-of-flight INS spectra of powder Nd$_2$Zr$_2$O$_7$ with incident neutron energy $E_i = 150$~meV and 400~meV at 5~K\@. These maps show the normalized scattering cross section $S(E,Q)$ where $E$ is the energy transfer and $Q$ is the scattering vector. While the high intensity around $E=0$ arises from elastic scattering, the scattering of phonons gives rise to a $Q$ dependent intensity which increase with increasing $Q$. In addition to these features we can clearly see three strong dispersionless excitations at low-$Q$ around 23.4~meV, 35.0~meV and 106.2~meV in the spectrum with $E_i=150$ meV [Fig.~\ref{fig:cf1}(a)] and two weaker ones near 240~meV and 310~meV (with even weaker levels in between these energies) in the spectrum with $E_i=400$ meV [Fig.~\ref{fig:cf1}(b)]. The $Q$-dependent integrated intensity between 105.7~meV and 106.7~meV follows the magnetic form factor $F^2(Q)$ of Nd$^{3+}$ \cite{formILL} as shown in Fig.~\ref{fig:cfform} and thus suggests that those excitations in the INS spectra result from the scattering of single-ion CEF transitions. 

\begin{figure}
\includegraphics[width=3.2in, keepaspectratio]{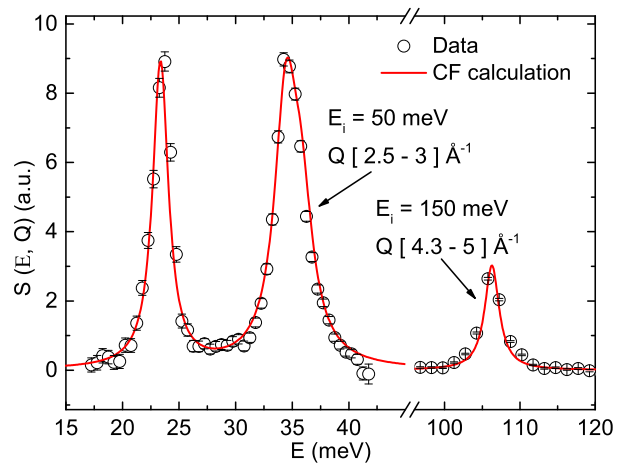}
\caption {(Color online) Fitted inelastic neutron spectrum  of the four transitions within the ground state multiplet $^4I_{9/2}$ at 5 K. The data for the three lower crystal field levels are from the dataset with incident neutron energy $E_i=50$~meV integrated $Q$ over the range [2.5--3]~{\AA $^{-1}$}. The data for the crystal field excitation at 106.2~meV are from the dataset with $E_i=150$~meV integrated $Q$ range [4.3--5]~{\AA $^{-1}$}.}
\label{fig:cf2}
\end{figure}

For the Nd$^{3+}$ ion, the Hund's-rule ground state (GS) multiplet is $^4I_{9/2}$ and the first excited multiplet, $^4I_{11/2}$, is typically 250 meV above it \cite{Boothroyd1992,bookliu}. Furthermore in pyrochlores the CEF splitting of the GS multiplet of rare earth ions is normally $\sim100$~meV \cite{cfhto,cfpso,nmpzo,cfttotso,Princep2015}. Therefore, we assign the three excitations below 200~meV to transitions within the GS multiplet and the two above 200~meV to intermultiplet transitions. According to Kramers theorem, the CEF interaction should split the GS multiplet $^4I_{9/2}$ of Nd$^{3+}$ into five doublets of $|\pm m_J\rangle$ type and thus there ought to be four excitations in the INS spectra at base temperature, corresponding to transitions from the GS doublet to the four excited doublet states. Although only three excitations within the GS multiplet at 23.4~meV, 35.0~meV and 106.2~meV are apparent, a closer inspection of the INS data (Fig.~\ref{fig:cf2}) reveals that the excitation at 35.0 meV is rather broad (compared to the instrument resolution function) which could be due to two unresolved excitations from two closely situated CEF levels (the so-called quasi-quartet). Similar unresolved excitations near 35.0~meV were also observed in INS data from a thermal neutron triple axis spectrometer in Ref.~\cite{nzoarxiv}. Two closely spaced CEF levels near 35.0~meV were also inferred by analyzing susceptibility and heat capacity data in Ref.~\cite{nzoprb}.

For a quantitative analysis of INS data we used crystal field model. In order to account for the mixing of the GS multiplet with the higher multiplets which is necessary for the situation with a large splitting of the GS multiplet (comparable with the energy separation of the first excited multiplet), we use tensor operators for the CEF Hamiltonian instead of Stevens' operator equivalents \cite{cfbook}. In the fcc pyrochlore structure of Nd$_2$Zr$_2$O$_7$ the Nd$^{3+}$ ions  are subjected to a crystal electric field with $D_{3d}$ symmetry created by the eight neighboring oxygen ions, and for the $z$ axis along the local cubic $\langle$111$\rangle$ direction the CEF Hamiltonian is given by \cite{cfbook},
\begin{equation}
\begin{split}
H_{\rm{CEF}}= & B_0^2C_0^2+B_0^4C_0^4+B_3^4(C_{-3}^4+C_3^4)\\
             & +B_0^6C_0^6+B_3^6(C_{-3}^6+C_3^6)+B_6^6(C_{-6}^6+C_6^6)
\end{split}
\end{equation}
where $B_q^k$ and $C_q^k$ are the crystal field parameters and the tensor operators, respectively. 

We used the intermediate-coupling free ion basis states for diagonalizing the $H_{\rm CEF}$. In a 4$f^n$ system, the static electric repulsion between the localized electrons splits the 4$f^n$ configuration into Russell-Saunders ($LS$-coupling) terms and the spin-orbital interaction mixes the $LS$-coupling terms with the same $J$ \cite{Boothroyd1992,bookliu}. The intermediate-coupling wave function is formed by a linear combination of $LS$-coupled states of the same $J$ with the Hund's-rule ground state as the dominate term. Because of the very large number of terms, it is not easy to handle all of them, so we only included the 98 intermediate coupling basis states from the first 12 multiplets below 2.2~eV in our calculation. 

\begin{table}
\caption{\label{tab:cf} Observed  and calculated  crystal-field transition energies ($E$) and integrated intensities ($I$) within the ground state multiplet $^4I_{9/2}$ of N$_2$Zr$_2$O$_7$ at 5~K\@. The $I$ is relative with respect to the highest peak observed.} 
\begin{ruledtabular}
\begin{tabular}{ccccc}
Levels & $E_{obs}$ (meV)& $E_{cal}$ (meV)&$I_{obs}$&$I_{cal}$\\
\hline
$\Gamma_{56}^+$ &0 & 0 & - & 2.5\\
$\Gamma_4^+$ & $23.4(2)$ & 23.36 & $0.58(5)$ & 0.558\\
$\Gamma_{56}^+$ & $34.4(4)$ & 34.44 & $\rceil  \hspace{3pt}$ 1 &0.655\\
$\Gamma_4^+$ & $35.7(4)$ & 35.81 & $\rfloor \hspace{10pt}$ &0.345\\
$\Gamma_4^+$ & $106.2(5)$ & 106.28 & $0.60(8)$ & 0.525\\
\end{tabular}
\end{ruledtabular}
\end{table}

INS measures the powder averaged unpolarized neutron inelastic scattering double-differential cross section given by \cite{Squires}
\begin{equation}
\begin{split}
\frac{d^2\sigma }{d\Omega dE'} = & \left(\gamma r_0\right)^2\frac{k'}{k}\sum_{\alpha \beta }\left(\delta _{\alpha \beta} -\hat{\kappa} _\alpha \hat{\kappa} _\beta \right)\sum_{\lambda \lambda '}p_\lambda\\
 &\times\left\langle\lambda \left|Q_\alpha ^\dagger \right|\lambda '\right\rangle\left\langle\lambda '\left|Q_\beta \right|\lambda \right\rangle\delta \left( E_\lambda -E_{\lambda'}+\hbar\omega \right),
\end{split}
\end{equation}
where $Q_\alpha$ is effectively the Fourier transform of the magnetization. For the calculation of the intensity of the transitions within the GS multiplet, we used the dipolar approximation allowing the above expression to be rewritten as   
\begin{equation}
\frac{d^2\sigma }{d\Omega dE'}=c F^2(Q)\frac{k'}{k} \sum_{\alpha}\sum_{\lambda \lambda '}p_\lambda \left|\langle\lambda '\left|J_\alpha \right|\lambda \rangle \right|^2 L_{\lambda \lambda '}
\end{equation}
where $c$ is a constant, $F^2(Q)$ is the magnetic form factor, $k$ and $k'$ are the moduli of the incident and scattered wave vectors,  $|\lambda '\rangle$ and $|\lambda\rangle$ are the initial and final eigenfunctions, $J_\alpha$ is the $x$, $y$ or $z$ component of the total angular momentum operator, and $L_{\lambda \lambda '}$ is the Lorentzian function describing the line shape of the excitation.

\begin{figure}
\includegraphics[width=3.2in, keepaspectratio]{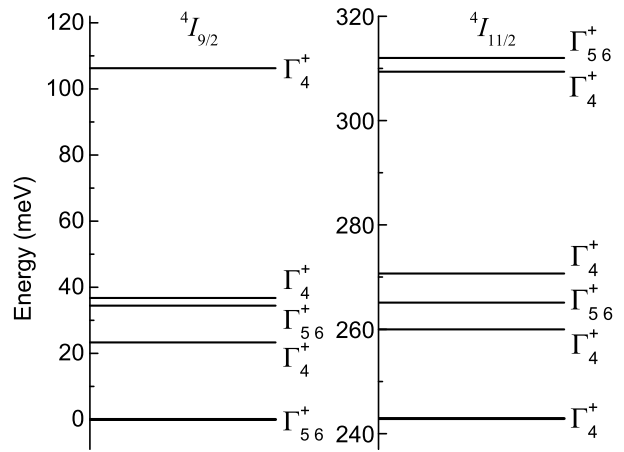}
\caption {(Color online) Crystal field energy schemes for (a) the ground-state multiplet $^4I_{9/2}$ and (b) the first excited multiplet $^4I_{11/2}$ corresponding to the crystal field parameters obtained from the analysis of INS data. $\Gamma$ shows the irreducible representation that the corresponding CEF state transforms as.}
\label{fig:cf3}
\end{figure}

The refinement of the INS spectra of the CEF transitions within the GS multiplet was performed using the program SPECTRE \cite{SPECTRE}, which was recently used successfully for analyzing the INS data of Pr$_2$Sn$_2$O$_7$ \cite{cfpso} and Tb$_2$Ti$_2$O$_7$ \cite{Princep2015}. The energy values and the integrated intensities of the levels were obtained by fitting the peaks in the spectrum with Lorentzian functions. As the two overlapping excitations near 35.0~meV are not resolved in our data, we used the combined intensity obtained by fitting them as a single peak. On the other hand, to get their approximate positions we fitted the peak with two Lorentz functions of same area and width. The low-$Q$ data were used where phonon scattering is weak and the La$_2$Zr$_2$O$_7$ data were used to identify phonons and to provide the non-magnetic background that was subtracted from the data. The CEF parameters in Ref.~\cite{nzoprb} were used as the starting parameter for the least-square fitting but scaled overall to match the calculated energy levels with the measured ones. We also tested the case where the two overlapping levels were exchanged.

For the best fit the CEF parameters are $B^2_0 = 49.2$~meV, $B^4_0 = 408.9 $~meV, $B^4_3 = 121.6 $~meV, $B^6_0 = 148.1 $~meV, $B^6_3 = -98.0 $~meV, and $B^6_6 = 139.1$~meV yielding standard normalized goodness-of-fit parameter $\chi^2=0.34$.  The fitting details are shown in Table~\ref{tab:cf} and the fitted spectrum is shown in Fig.~\ref{fig:cf2}. As listed in Table~\ref{tab:cf} these CEF parameters correspond to five doublets at 0, 23.4~meV, 34.4~meV, 35.8~meV and 106.3~meV. In order to compare with other related reports, we converted these parameters into the Steven's formalism by using the relation $D^q_k=\Lambda \lambda^q_k B^q_k$ ($\Lambda=\alpha_J,\ \beta_J \ {\rm and}\  \gamma_J$ listed in Ref.~\cite{hutchings} and $\lambda^q_k$ are listed in Ref.~\cite{bookliu}). After the transformation we get $D^2_0 = -0.158$~meV, $D^4_0 = -0.0149 $~meV, $D^4_3 = 0.105 $~meV, $D^6_0 = -0.00035 $~meV, $D^6_3 = -0.0048 $~meV, and $D^6_6 = -0.005$~meV. These CEF parameters are comparable with those obtained for Nd$_2$Zr$_2$O$_7$ from the CEF analysis of susceptibility data in Ref.~\cite{nzoprb}. However our results differ significantly with those in Ref.~\cite{nzoarxiv}, where only the two lowest excitations were measured using a triple axis spectrometer and were analyzed within the ground state multiplet using Stevens' operator equivalents. Our analysis is also consistent with that of Pr$_2$Sn$_2$O$_7$ \cite{cfpso}.

\begin{figure}
\includegraphics[width=3.2in, keepaspectratio]{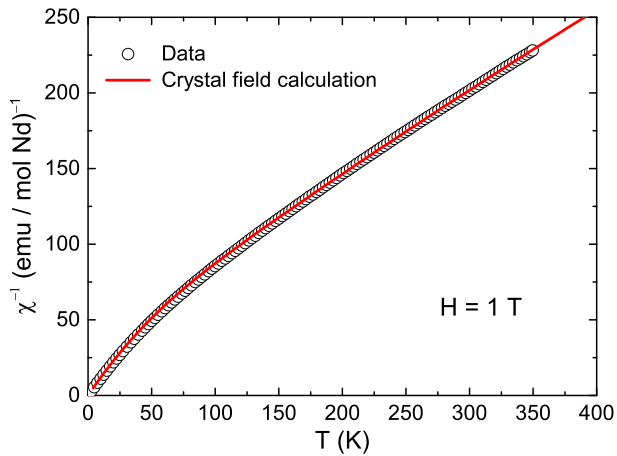}
\caption {(Color online) Inverse magnetic susceptibility $\chi^{-1}$ versus temperature $T$ of powder Nd$_2$Zr$_2$O$_7$ measured in a field of 1~T\@. The solid curve is the crystal field susceptibility corresponding to the crystal field parameters obtained from the analysis of INS data.}
\label{fig:cf4}
\end{figure}

The refined CEF energy scheme of the GS multiplet $^4I_{9/2}$ is shown in Fig.~\ref{fig:cf3} together with the calculated scheme for the first excited multiplet $^4I_{11/2}$ which also matches the experimental INS data well (not shown). The wavefunctions of the GS doublet and the first excited doublets within the GS multiplet $^4I_{9/2}$ are found to be
\begin{subequations}
\begin{equation}
\begin{split}
\Gamma_{56}^+  = & 0.899|^4I_{9/2},\pm 9/2 \rangle \mp 0.252|^4I_{9/2}, \pm 3/2 \rangle \\ 
                 &+ 0.330|^4I_{9/2}, \mp 3/2 \rangle \mp 0.112|^4I_{11/2},\pm 9/2 \rangle  \\ 
\end{split}
\end{equation}
and
\begin{equation}
\begin{split}
\Gamma_4^+   =  & 0.149|^4I_{9/2}, \pm 7/2 \rangle + 0.743|^4I_{9/2},\mp 5/2 \rangle  \\ 
               & \mp 0.643|^4I_{9/2}, \pm 1/2 \rangle \pm 0.056|^4I_{11/2},\pm 7/2 \rangle. 
\end{split}
\end{equation}
\label{eq:wavefunctions}
\end{subequations}
As we can see from Eq.~(\ref{eq:wavefunctions}a), there is a large mixing of $|^4I_{9/2},\pm 9/2 \rangle$ with $|^4I_{9/2},m_J\neq\pm9/2 \rangle$ terms in the ground state as well as a small mixing with $^4I_{11/2}$ leading to reduction in the moment of Nd$^{3+}$. The ground state moment calculated from Eq.~(\ref{eq:wavefunctions}a) is $2.65\,\mu_{\rm B}$ with $g_{\rm zz}\approx 5.30$ and $g_{\bot}=0$ which indicates an Ising anisotropy. Moreover the Ising anisotropy can be regarded as very strong when considering the large first excitation energy 23.4~meV. These values agree well with the magnetic data in Sec.~\ref{Magnetic}. However our $g_{\rm zz}$ is different from the value of 4.793 in Ref.~\cite{nzoprb} and the value of 4.3 in Ref.~\cite{nzoarxiv}. The difference in $g_{\rm zz}$ results probably from different sets of CEF parameters obtained in Refs.~\cite{nzoprb,nzoarxiv}. Since we have accessed the higher energy excitations we believe that our CEF parameters are more accurate. Furthermore, the ground state is exactly diploar-octuploar doublet which transforms as $\Gamma^+_{56}$ expected when $D_0^2<0$ and dominates the other terms \cite{dioctu}. The magnetic susceptibility calculated according to the CEF parameters is shown in Fig.~\ref{fig:cf4}. A very good agreement between the experimental data and the calculation supports the validity of our analysis of the INS data and the extracted CEF parameters.

\section{\label{Conclusion} Summary and Conclusions}

We have investigated the magnetic structure and crystal field states of the pyrocholre compound Nd$_2$Zr$_2$O$_7$. The high-resolution synchrotron x-ray powder diffraction reveals the pyrochlore structure ($Fd\bar{3}m$) of Nd$_2$Zr$_2$O$_7$ without any observable oxygen deficiency or Nd/Zr site mixing. A positive $\theta_{\rm p}$ in $\chi(T)$ indicates a ferromagnetic coupling between the Nd$^{3+}$ moments although the effective nearest neighbour interaction is antiferromagnetic leading to a long range antiferromagnetic order. The neutron diffraction data reveals an all-in/all-out magnetic structure below $T_{\rm N}\approx 0.4~$K with an ordered moment $1.26(2)\, \mu_{\rm B}$/Nd at 0.1~K\@. The ground state moment has Ising anisotropy and is estimated to be $\sim 2.65\,\mu_{\rm B}$/Nd giving an effective $S = 1/2$ doublet with $g_{\rm zz} = 5.30(6)$ and $g_{\bot}=0$, which are deduced from the magnetization and INS data. The crystal field eigenvalues and eigenvectors have been determined by analyzing inelastic neutron scattering data which confirm the strong local $\langle111\rangle$ Ising anisotropy and the dipolar-octupolar nature of Nd$^{3+}$ moments with a large separation of 23.4~meV between the ground state doublet and the first excited doublet.

We see a strongly reduced ordered state moment of $1.26(2)\, \mu_{\rm B}$/Nd compared to the estimated Ising moment of $ 2.65\,\mu_{\rm B}$/Nd indicating persistent quantum fluctuations deep into the ordered phase. According to the phase diagram of Ising pyrochlore \cite{dsim2004} the AIAO antiferromagnetic state is a stable state without any kind of frustration in the ordered state. As such the reduction in moment could not be understood to result on account of frustration. We therefore suspect that the strong quantum fluctuations and reduction of moment could be an attribute of octupolar tensor component of ground state doublet \cite{dioctu}. Because of the octupolar term the Nd$_2$Zr$_2$O$_7$ may not behave strictly like a dipolar system and this non-Ising term can cause quantum fluctuations. Further investigations are desired to access the attributes arising from dipolar octupolar nature of ground state Kramers doublet and associated double monopole dynamics.

\acknowledgements
We thank A. T. M. N. Islam for his help in sample preparation, B. Klemke for his assistance in measurements using PPMS, F. Yokaichiya for his help in refining XRD data, A. T. Boothroyd for help on crystal field analysis and Y.-P. Huang and M. Hermele for helpful discussions on the related theory.  We acknowledge Helmholtz Gemeinschaft for funding via the Helmholtz Virtual Institute (Project No. VH-VI-521).

\end{document}